\documentclass[a4paper,UKenglish]{lipics}
\usepackage{microtype}%if unwanted, comment out or use option "draft"

\title{Realistic cost for the model of coherent computing}
\titlerunning{Realistic cost for the model of coherent computing} %optional, in case that the title is too long; the running title should fit into the top page column

\author[1]{Akira SaiToh}
\affil[1]{National Institute of Informatics\\
  2-1-2 Hitotsubashi, Chiyoda, Tokyo 101-8430, Japan\\
  \texttt{akirasaitoh@nii.ac.jp}}
\authorrunning{A. SaiToh} %mandatory. First: Use abbreviated first/middle names. Second (only in severe cases): Use first author plus 'et. al.'

\Copyright[by]{A. SaiToh}%mandatory. Default is "by";  http://creativecommons.org/licenses/by/3.0/

% mandatory: Please choose ACM 1998 classifications from http://www.acm.org/about/class/ccs98-html . E.g., cite as "F.1.1 Models of Computation". 
% mandatory: Please provide 1-5 keywords
\serieslogo{}%please provide filename (without suffix)
\volumeinfo%(easychair interface)
  {Billy Editor, Bill Editors}% editors
  {2}% number of editors: 1, 2, ....
  {Conference title on which this volume is based on}% event
  {1}% volume
  {1}% issue
  {1}% starting page number
\EventShortName{}
\DOI{10.4230/LIPIcs.xxx.yyy.p}% to be completed by the volume editor
\theoremstyle{plain}
\newtheorem{proposition}{Proposition}
\begin{document}

\maketitle

\begin{abstract}
For the model of so-called coherent computing recently proposed by Yamamoto {\em et al.}
[Y. Yamamoto {\em et al.}, New Gen. Comput. 30 (2012) 327-355], a theoretical analysis
of the success probability is given. Although it was claimed as their prospect that the 
Ising spin configuration problem would be efficiently solvable in the model, here it is shown that the
probability of finding a desired spin configuration decreases exponentially in the number of spins
for certain hard instances. The model is thus physically unfeasible for solving the problem within a
polynomial cost.
\end{abstract}

\section{Introduction}
It has been of long-standing interest to study the ability of analog computing systems to
solve computationally difficult problems \cite{R93,ERN08}. It is recently of growing interest to
investigate the power of quantum adiabatic time evolution in this direction \cite{F01}.
Nevertheless, it has been commonly believed, with strong theoretical and numerical evidences,
that a desired solution should not be obtained with a sufficiently large probability within
polynomial time owing to the exponential decrease in the energy gap between desired and undesired
eigenstates during an adiabatic change of Hamiltonians \cite{D01,Z05,Z06,A08,HY11,F12}.

Recently, Yamamoto {\em et al.} wrote a series of papers \cite{UTY11,TUY12,YTU12}
on their model\textemdash so called the coherence computing model\textemdash of an injection-locked slave
laser network, which uses quantum states to some extent in contrast to conventional classical optical
computing models \cite{SMDR07,OSC2008}.
It was claimed to be promising in solving the Ising spin configuration problem \cite{B82} and those
polynomial-time reducible to this problem faster than known conventional models.

The Ising spin configuration problem has been well-known as a typical NP-hard problem described by
an Ising-type Hamiltonian \cite{B82}. A typical description is as follows.
\begin{quote}
{\em Ising spin configuration problem:}
Given a graph $G=(V,E)$ with set $V$ of vertices and set $E$ of edges, and
weighting functions $J:E\rightarrow\{0,\pm 1\}$ and $B:V\rightarrow\{0,\pm 1\}$, find
the minimum eigenvalue $\lambda_{\rm g}$ of the Hamiltonian
$H=\sum_{(ij)\in E}J_{ij}\sigma_{z,i}\sigma_{z,j}+\sum_{i\in V}B_i\sigma_{z,i}$. Here,
$\sigma_{z,i}$ is the Pauli Z operator acting on the space of the $i$th spin (there are $n=|V|$
spin-1/2's).
\end{quote}
In an intuitive point of view, the problem is difficult in the sense that the number of given
parameters grows quadratically while the number of eigenvalues including multiplicity grows
exponentially. Although the Hamiltonian is diagonal in the Z basis, writing it in the matrix
form itself takes exponential time. Hereafter, we employ $n$ for representing the input
length of an instance although, precisely speaking, the bit length of an encoded instance is
$O(n^2)$. We do not go into the controversy on the definition of the input length \cite{ONeil09}.
As for known results on the complexity of the problem, it becomes P in case the graph is a
planer graph and $B_i=0~~\forall i$ (see Ref.~\cite{Is00}); for nonplaner graphs, it is in
general NP-hard, and it is so under many different conditions \cite{Is00}. In addition, a planer graph
together with nonzero $B_i$'s also makes the problem NP-hard \cite{B82}. It is also worthwhile to mention
that the typical value of $\lambda_{\rm g}$ is $c_{\rm g}n$ with coefficient $c_{\rm g}$ (so-called the ground-state energy
density) typically between $-2$ and $-1/2$ when the values of $J_{ij}$ are chosen in a certain random
manner and $B_i$ are set to zero \cite{VT77,Kirkpatrick77,MB80,Derrida80,Derrida81,Simone95,Andreanov04,Boettcher10}
($c_{\rm g}$ is between $-1.5$ and $-1$ when the graph is a ladder and $J_{ij}$ and $B_i$ are randomly
chosen from $\{\pm1\}$ \cite{Kadowaki95}). Furthermore, it should be mentioned that the distribution
of eigenenergies of $H$ (namely, the envelope of the multiplicity of eigenenergies with a normalization) is a
normal distribution with mean zero and standard deviation proportional to $\sqrt{n}$ in the random energy
model \cite{Derrida80,Derrida85,Andreanov04}. Here, the important observation is that the standard deviation
increases with $n$ in spite of the exponentially increasing number of spin configurations.

Let us also introduce the NP-complete variant of the Ising spin configuration problem as follows.
\begin{quote}
{\em NPC Ising spin configuration problem:}\\
{\em Instance:}
Positive integer $n$, integer $K$, and parameters $J_{ij}\in\{0,\pm 1\}$ ($i < j$)
and $B_i\in\{0,\pm 1\}$ for integers $0\le i,j \le n-1$.\\
{\em Question:}
Is there an eigenvalue $\lambda$ of the Hamiltonian $H=\sum_{i,j=0; i < j}^{n-1} J_{ij}\sigma_{z,i}\sigma_{z,j}
+\sum_{i=0}^{n-1}B_i\sigma_{z,i}$ such that $\lambda < K$ ?
\end{quote}
This is the problem we are going to investigate in this contribution as for its computational difficulty
under the coherent computing model.

\begin{figure}[ptb]
\begin{center}
\resizebox{0.82\textwidth}{!}{\includegraphics{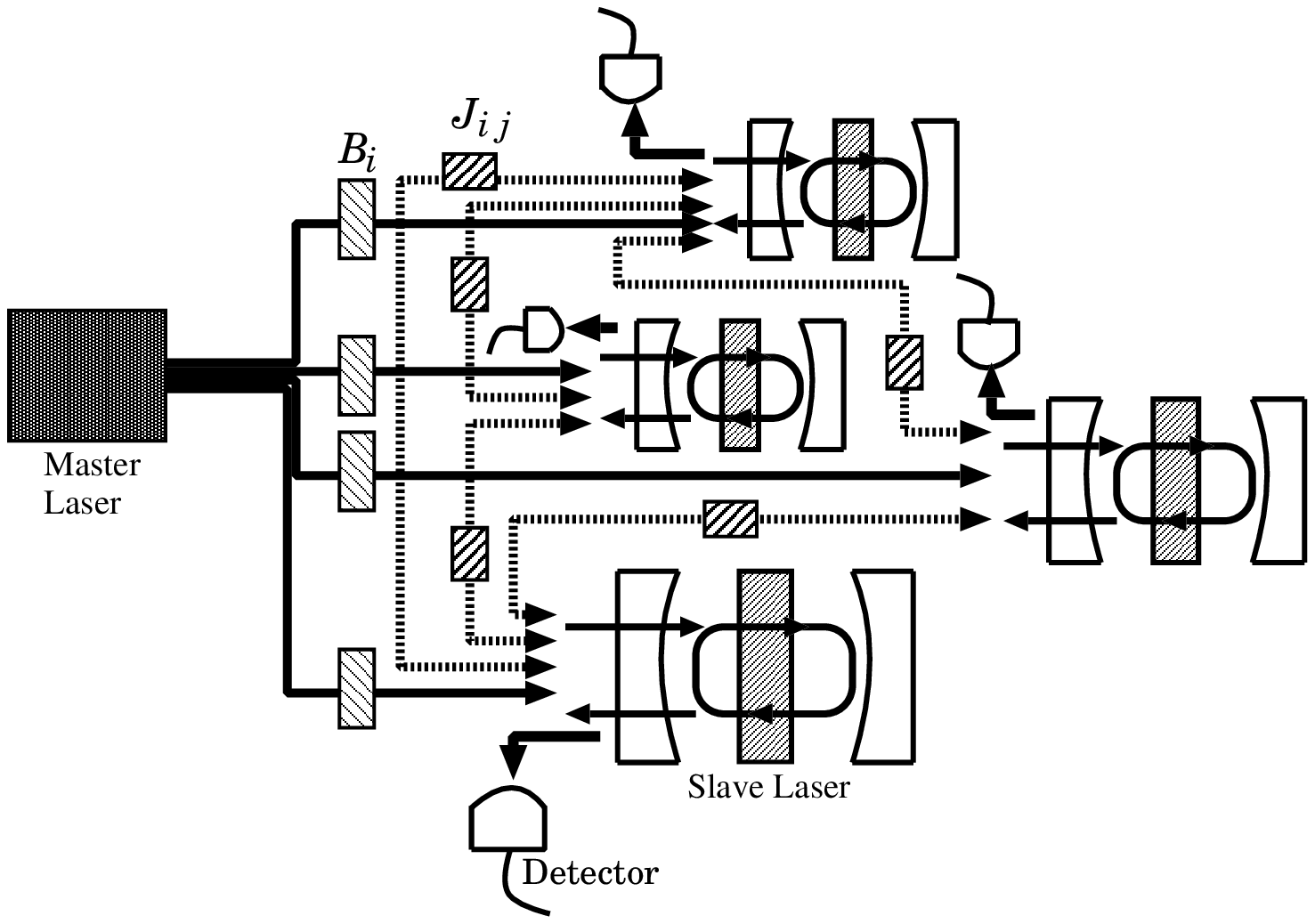}}
\caption{\label{figlasernetwork}
Schematic description of the coherent computing model. See the text for how $J_{ij}$ and $B_i$
are realized by optical instruments.}
\end{center}
\end{figure}

Let us now briefly look into Yamamoto {\em et al.}'s coherent computing model \cite{UTY11,TUY12,YTU12} which
is schematically depicted as Fig.~\ref{figlasernetwork}.
It has one master laser whose output is split into $n$ paths and injected to $n$ slave lasers.
Each slave laser is initially locked to the superposed state $(|R\rangle_i+|L\rangle_i)$
where $|R\rangle$ and $|L\rangle$ are the right and left circular polarized states
(see, {\em e.g.}, Refs.~\cite{H67,HY84} for physics of the injection-locked laser system).
The initial state of the $n$ slave lasers is therefore $\bigotimes_{i=0}^{n-1}(|R\rangle_i+|L\rangle_i)$.
The laser network is a macroscopic system; thus initially it holds many photons in this same state.
The computational basis is set to $\{|R\rangle,|L\rangle\}^n$ and $\sigma_z$ is written as
$|R\rangle\langle R|-|L\rangle\langle L|$.
The $i$th slave laser and the $j$th slave laser are connected for nonzero $J_{ij}$. At time
$t=0$, they mutually inject a small amount of horizontally polarized signal via an attenuator,
a phase shifter, and a horizontal linear polarizer, which determine the amplitude attenuation
coefficient that is regarded as $J_{ij}$. Among the three instruments, the attenuator's transmission
coefficient controls $|J_{ij}|$ and the other instruments controls ${\rm sgn} J_{ij}$.
In addition, a small amount of injection of horizontally polarized signal is also made from the
master laser to each slave laser at $t=0$. This amount corresponds to $B_i$ for the $i$th slave laser.
It is controlled by the combination of a half-wave plate and a quarter wave plate.
For more details of implementation of the coefficients, see section 7 of Utsunomiya {\em et al.}~\cite{UTY11}.

Then one waits for a small time duration $t_{\rm st}$ to let the system evolve. Laser modes satisfying
the matching condition with the above-mentioned setting grow rapidly and other modes are suppressed.
For $t>t_{\rm st}$, the system is thought to be in a steady state. Then for each slave laser its output
is guided to a polarization beam splitter and the right and the left polarization components are separately
detected by photodetectors. By a majority vote of photon number counting, the computational result of each
slave laser, $|a\rangle_i\in\{|R\rangle,|L\rangle\}$, is retrieved. The steady state
$|a\rangle_0\cdots|a\rangle_{n-1}$ is thus determined. Once this is determined,
it takes only polynomial time to calculate the corresponding eigenvalue since there are
only $O(n^2)$ terms in the Hamiltonian (here, we do not use its matrix form).

Thus, in short, the state starts from $(|R\rangle+|L\rangle)^{\otimes n}$ and eventually reaches a
steady state representing a configuration that corresponds to the minimum energy of the given Hamiltonian.
Yamamoto {\em et al.} \cite{UTY11,TUY12,YTU12} employed rate equations involving several factors
characterizing each oscillator and connections with other oscillators to analyze photon numbers of the right
and left polarization components for each slave laser; they concluded that the system reaches a steady state
within 10 nano seconds without obvious dependence on $n$.

It has been unknown so far if the coherent computing model is a valid computer model in view of
a rigid and fair description of computational costs. Conventional analog computing models do not
solve NP-hard problems within a polynomial cost; they require either exponentially long convergence time
or exponentially fine accuracy \cite{Aa05}. Thus it should be natural to be skeptical against the power
of the coherent computing model. In this contribution, we investigate the signal per noise ratio in the
output of the coherent computer when the NPC Ising spin configuration problem is handled.
%%%
We will reach the fact that for certain hard instances, the relative signal intensity corresponding to solutions
is bounded above by a function decreasing exponentially in $n$. This is because the number of modes that
are possibly locked in the laser network increases rapidly in $n$ owing to the fact that the locking range of
the laser network does not shrink as $n$ grows considering imperfectness of optical instruments.

The analysis of computational difficulty is described in Sec.~\ref{secanalysis}. The result is
discussed in Sec.~\ref{secdiscussion} and summarized in Sec.~\ref{secconclusion}.

\section{Computational difficulty in the coherent computing model}\label{secanalysis}
The coherent computing model illustrated in Fig.~\ref{figlasernetwork} was so far analyzed
by Utsunomiya {\em et al.} \cite{UTY11,TUY12,YTU12} on the basis of the assumption that given
coefficients $J_{ij}$ and $B_{i}$ are exactly implemented by optical instruments although fluctuations
and quantum noise in the system were considered in their analyses of time evolutions using rate equations, which led
to a quite ideal convergence taking only 10 nano seconds.

\setcounter{footnote}{1}
Here, we assume that individual optical instruments are imperfect\footnotemark[1] so that there are errors in
$J_{ij}$ and $B_{i}$, which are due to calibration errors and/or thermal fluctuations.
\footnotetext[1]{It is a common case that each optical instrument has a few permil uncertainty
in the calibration of each property (see Ref.~\cite{sp250}).
In addition, there is a quantum limit in any classical instrument \cite{Cl10,La04}
so that a manufacturing error and a manipulation error cannot be made arbitrarily small.}
Then the following proposition is achieved.
\begin{proposition}\label{prop1}
Consider the NPC Ising spin configuration problem. 
Suppose calibration errors and/or thermal fluctuations of optical instruments cause
nonzero physical deviations,\footnotemark[1]
$\;\varepsilon_{ij}\in{\bf R}$ for nonzero $J_{ij}$ and $\kappa_i\in{\bf R}$ for nonzero $B_{i}$.
We assume that $\varepsilon_{ij}$ are {\em i.i.d.} random variables with mean zero and a certain standard deviation
$\sigma_\varepsilon$ and $\kappa_i$ are {\em i.i.d.} random variables with mean zero and a certain standard deviation
$\sigma_\kappa$. Then, for large $n$, there exist YES instances such that the probability to obtain a spin
configuration corresponding to one of $\lambda{\textrm 's}<K$ using the coherent computer is
$\le {\rm poly}(n)2^{-n}$.
\end{proposition}
The proof is given as below.\\
~\\
{\bf Proof of Proposition \ref{prop1}}\\
Here we consider instances generated in the way that $J_{ij}$'s and $B_i$'s are independent uniformly distributed
random variables with values in $\{0, \pm 1\}$. Since a problem instance is a given data set, the standard deviation
for $J_{ij}$ and that for $B_i$ intrinsic to the problem instance itself are not of our concern. We only consider
physical deviations as errors.

As the model is a sort of a bulk model (there are many photons), it is convenient to consider
populations of individual configurations. Let $P_{\lambda,l_\lambda}(t)$ be the population of each eigenstate
$|\varphi_{\lambda,l_\lambda}\rangle$ ($l_\lambda\in\{0, \ldots, d_\lambda-1\}$) corresponding to eigenenergy $\lambda$
of the Hamiltonian (the Hamiltonian is specified by the problem instance), where $t$ stands for time and $d_\lambda$ is
the multiplicity of $\lambda$. We also introduce $P_\lambda(t)=\sum_{l_\lambda=0}^{d_\lambda-1} P_{\lambda,l_\lambda}(t)$.
It should be kept in mind that we do not start from the thermal distribution; for the initial state,
we have identical copies of $\sum_\lambda\sum_{l_\lambda}|\varphi_{\lambda,l_\lambda}\rangle=(|R\rangle+|L\rangle)^{\otimes n}$.
In the present setting, the random-energy model \cite{Derrida80,Derrida85} is valid\footnote{
Let us pick up a certain configuration $|\varphi\rangle$. Suppose, by applying $m$ bit flips,
its energy changes by $\Delta E(\varphi\overset{m}{\mapsto}\varphi')$ with $|\varphi'\rangle$ a resultant configuration.
This process should obey the random energy change and hence for large $m$,
$\Delta E(\varphi\overset{m}{\mapsto}\varphi')$ should obey the normal distribution with mean zero and a standard
deviation proportional to $\sqrt{m}$ by the central limit theorem (in regard with a sum of random variables).
In addition, the most typical number of bit flips is $n/2$ when we generate all other configurations
from $|\varphi\rangle$. Typical bit flips generate a dominant number of configurations.
Thus the distribution of energies is approximated by the normal distribution with mean zero and a standard
deviation proportional to $\sqrt{n}$. In this way, we have just obtained the distribution of energies
in the random-energy model.
} and hence, for large $n$, with an appropriate scaling factor $M$, one can write
$P_\lambda(0)=M\mathcal{N}(0, \sigma_\lambda^2)$ with $\sigma_\lambda=\Theta(\sqrt{n})$ where $\mathcal{N}(\mu, \sigma^2)$ is
the density function of the normal distribution with mean $\mu$ and standard deviation $\sigma$.
Here, we have $M = 2^nP_{\lambda_{\rm g},0}(0)$ with $\lambda_{\rm g}$ the ground state energy 
because the initial population is same for all the configurations.

Let us denote the set of solution states (spin configurations corresponding to
$\lambda{\textrm 's}<K$) as $Y$.
The total population of solution states at $t$ is given by $P_{Y}(t)=\sum_{\lambda < K}P_\lambda(t)$.
Similarly, the total population of nonsolution states is given by $P_{X}(t)=\sum_{\lambda\ge K}P_\lambda(t)$; here,
$X=\{|\varphi_{\lambda,l_\lambda}\rangle\;|\;\lambda\ge K\}$. Ideally, only $|\varphi_{\lambda,l_\lambda}\rangle{\textrm 's}\in Y$
will enjoy population enhancement by mode selections. However, there exists $v \ge K$ such that
$P_\lambda(t>t_{\rm st}) \gg 0$ for $\lambda \le v$. This is because the matching condition is imperfect in reality;
the locking range is broader than the ideal range considering errors in optical
instruments.\footnote{See, {\em e.g.}, Ref.~\cite{KK81} for an experimental gain curve.}
Let us write $P_{Z}(t)=\sum_{K \le \lambda \le v}P_\lambda(t)$; here, $Z=\{|\varphi_{\lambda,l_\lambda}\rangle\;|\;K \le \lambda \le v\}$.

By assumption, we are considering physical deviations (including calibration errors and thermal fluctuations),
$\varepsilon_{ij}$ for nonzero $J_{ij}$ and $\kappa_i$ for nonzero $B_i$.
The Hamiltonian implemented on the laser network is written as
$\widetilde{H}=\sum_{i<j|J_{ij}\not = 0}(J_{ij}+\varepsilon_{ij})\sigma_{z,i}\sigma_{z,j}
+\sum_{i|B_i\not = 0}(B_i+\kappa_i)\sigma_{z,i}$.
This suggests that $v=K+K'(n)$
with $K'(n)\simeq\sigma_\varepsilon \sqrt{n(n-1)/3}+\sigma_\kappa\sqrt{2n/3}$
by the central limit theorem in regard with a sum of random variables (see, {\em e.g.}, Refs.~\cite{Shiryaev,Klenke}),
considering the expected number of nonzero $J_{ij}$'s and that of nonzero $B_i$'s.
Therefore, $P_{Z}(0)=M\int_K^{K+K'(n)}\mathcal{N}(0, \sigma_\lambda^2){\rm d}\lambda$.

Let us write $H=H_J+H_B$ with $H_J=\sum_{i<j}J_{ij}\sigma_{z,i}\sigma_{z,j}$ and $H_B=\sum_iB_i\sigma_{z,i}$.
As we have mentioned, it is known \cite{VT77,Kirkpatrick77,MB80,Derrida80,Derrida81,Simone95,Andreanov04,Boettcher10}
that the ground state energy of $H_J$ is typically $c_{\rm g} n$ with $-2<c_{\rm g}<-1/2$.
Therefore, for any normalized vector $|v\rangle$ in the Hilbert space of the system of our concern,
$\langle v|H|v\rangle$ is typically bounded below by $-3n$. Thus, for typical instances we can choose
$K=K(n)$ with $-K(n)=O(n)$. Recall that $K'(n)=\Theta(n)$ and $\sigma_\lambda=\Theta(\sqrt{n})$.
We find that $\int_K^{K+K'(n)}\mathcal{N}(0, \sigma_\lambda^2){\rm d}\lambda
=\left[\frac{1}{2}{\rm erf}(\frac{\lambda}{\sqrt{2}\sigma_\lambda})\right]_{K}^{K+K'(n)}$ is a monotonically increasing
function of $n$.
%%%This is simply because the derivative (as a function of n) is positive for any n > 0.
Hence, for a certain constant $b>0$, $P_{Z}(0)\ge b2^nP_{\lambda_{\rm g},0}(0)$. 

Let us assume that locked modes have equally enhanced intensities for $t>t_{\rm st}$.
This leads to the signal per noise ratio for $t>t_{\rm st}$:
$P_{Y}(t>t_{\rm st})/P_{Z}(t>t_{\rm st}) = P_{Y}(0)/P_{Z}(0)$.
(In case one can assume that only one of $|\varphi_{\lambda,l_\lambda}\rangle$'s in $Y\cup Z$ survives, the ratio of the
probability of finding $|\varphi_{\lambda,l_\lambda}\rangle$ originated from $Y$ and that of finding
$|\varphi_{\lambda,l_\lambda}\rangle$ originated from $Z$ at $t>t_{\rm st}$ is given by the same equation.)

Consider some typical instances for which $d_{\rm g}$ is small and is not clearly dependent on $n$
($d_{\rm g}$ is the multiplicity in the ground level). This is a typical situation because the multiplicity
of $\lambda$ obeys the distribution $\mathcal{N}(0,\sigma_\lambda^2)$ with $\sigma_\lambda=\Theta(\sqrt{n})$ in the
present setting, as we have explained.
It is always possible to choose\footnote{Recall that we are proving the {\em existence} of hard instances.}
the value of $K$ such that all $|\varphi_{\lambda,l_\lambda}\rangle\in Y$ are configurations with at most a
constant number of bits different from one of the ground states.
In this case, $P_{Y}(0)={\rm poly}(n)P_{\lambda_{\rm g},0}(0)$ and thus, for large $n$,
$P_{Y}(t>t_{\rm st})/P_{Z}(t>t_{\rm st})\le {\rm poly}(n) 2^{-n}$.
$\Box$

\begin{remark}\label{rem1}
It is trivial to find a similar proof for the existence of hard instances of the Ising spin
configuration problem for finding a ground level in the coherent computing model.
\end{remark}

By Proposition~\ref{prop1}, it is now easy to prove the following theorem.
\begin{theorem}\label{theo1}
There exists an instance of the NPC Ising spin configuration problem such that a decision
takes $\Omega(2^{n}/{\rm poly}(n))$ time in the coherent computing model when nonzero physical
deviations,\footnotemark[1] $\;\varepsilon_{ij}\in{\bf R}$ for nonzero $J_{ij}$ and
$\kappa_{i}\in{\bf R}$ for nonzero $B_{i}$, are considered. Here, $\varepsilon_{ij}$
($\kappa_{i}$) are assumed to be {\em i.i.d.} random variables with zero mean and a certain standard
deviation $\sigma_\varepsilon$ ($\sigma_\kappa$).
\end{theorem}
{\bf Proof of Theorem~\ref{theo1}}\\
By Proposition~\ref{prop1}, there exists an YES instance such that the probability $p_s$
for a single trial of coherent computing to find $\lambda<K$ is $\le{\rm poly}(n)2^{-n}$.
The success probability after $\tau$ trials is given by $1-(1-p_s)^\tau$. In order to make
this probability larger than a certain constant $c>0$, we need $\tau>\log(1-c)/\log(1-p_s)
= (\log\frac{1}{1-c})/[p_s+\mathcal{O}(p_s^2)]=\Omega(2^{n}/{\rm poly}(n))$.
$\Box$

\section{Discussion}\label{secdiscussion}
We have theoretically shown a weakness of the coherent computing model for the problem to examine the
existence of a suitably small (large negative) eigenvalue of an Ising spin glass Hamiltonian. As the
number $n$ of spins grows, the desired signal decreases exponentially for certain hard instances because
exponentially many undesired configurations obtain gains in a realistic setting.

Indeed, Yamamoto {\em et al.} made numerical simulations \cite{UTY11,TUY12,YTU12} to examine their
prospect that a desired configuration would be found efficiently in the coherent computing model.
But, in general, the following points should be taken into account whenever a computer simulation of
a physical system is performed.

First, in classical computing, exponentially fine accuracy is achievable by linearly increasing the
register size of a variable or an array size of combined variables. Nevertheless, in physical systems,
noise decreases as $\propto 1/\sqrt{T}$ with $T$ the number of trials or the number of identical systems
according to the well-known central limit theorem. In the field of quantum computing, this has been
well-studied in the context of NMR bulk-ensemble computation at room temperature which suffers from
exponential decrease of signal intensity corresponding to the computation result as the input size grows
(see, {\em e.g.}, \cite{KCL98,SK06}). In the coherent computing model, the ratio of the population of correct
configurations and that of wrong configurations at the steady state should not decrease in a super-polynomial
manner if the model were physically feasible for solving the problem efficiently.
%%%
So far, Yamamoto {\em et al.} reported \cite{UTY11,TUY12,YTU12} that each slave laser maintains a sufficiently large
discrepancy between the populations of $|R\rangle$ and $|L\rangle$ at the steady state for some instances with a small
number of spins ($n\le 10$), using a simulation based on rate equations. They also showed their simulation results for
$n=1000$ for a very restricted type of instances such that $J_{ij}$'s take the same value and $B_i$'s for odd
$i$ take the same value and so do for even $i$. Nevertheless, the populations (in other words, the joint probabilities)
of correct and wrong configurations and how they scale for large $n$ were not reported.
%%%
Recently, Wen \cite{Wen12} showed his simulation results for the case where the graph was a two-layer lattice
for $n$ up to $800$. Although it was reported that his simulations of the coherent computer found eigenvalues
lower than those found by a certain semidefinite programming method, the populations of correct and wrong configurations
were not shown. Thus, it is difficult to discuss the power of the coherent computing model on the basis of presently
known simulation results.

Second, the coefficients of a problem Hamiltonian cannot be implemented as they are, in reality. Seemingly
negligible errors in the coefficients might be crucial in complexity analyses for a large input size.
This point has not been considered in conventional simulation studies \cite{TUY12,YTU12,Wen12} of the coherent
computing model.
%%%
In the coherent computing model, nonzero $J_{ij}$'s and nonzero $B_i$'s in the Ising spin glass Hamiltonian should
accompany calibration errors and/or thermal fluctuations. In particular, optical instruments usually have
nonnegligible calibration errors \cite{sp250}. As we have written in the proof of Proposition~\ref{prop1},
a well-known application of the central limit theorem for the sum of random variables \cite{Shiryaev,Klenke}
indicates the important observation that the sum of such physical deviations is an increasing function of the number
of spins. This fact has led to our conclusion that the relative population of desired configurations decreases
exponentially in $n$ for certain hard instances.

The second point is also usually overlooked in computer simulations \cite{F01} of adiabatic quantum computing.
Discussions on the complexity of adiabatic time evolution are usually made as to how long time should be
spent in light of a minimum energy gap between the ground state and the nearest excited state during
adiabatically changing the Hamiltonian toward its final form. The coefficients in the starting and
the final Hamiltonians are quite often considered to be given accurate numbers \cite{F12}. Nevertheless, they
should have certain errors due to imperfect calibrations \cite{sp250} and/or fluctuations in reality,
as we have discussed.
%%%
The target state will not appear as a stable state if a nontarget state of the final Hamiltonian becomes a ground
state of the Hamiltonian owing to the errors. A real physical setup for adiabatic quantum computing should suffer
from the demand of considerably fine tuning of individual apparatus to implement desired coupling for large $n$.
So far, $n$ has not been very large in physical implementations \cite{S03,P08,J11} so that this problem has not been
significant. (In addition, even under the setting without error in Hamiltonian coefficients, adiabatic quantum
computing tends to suffer from exponentially decreasing energy gap when random instances of certain NP-hard problems
are tried, according to the numerical analysis by Farhi {\em et al}.\cite{F12})

A possible way to avoid very fine tuning is to use error correction schemes similar to those
for standard circuit-model quantum computing. There have been several studies on error correction
codes \cite{J06} and dynamical decoupling \cite{L08,Qu12} in the context of adiabatic quantum computing.
It is of interest if similar schemes apply to the coherent computing model.
As for error correction codes, each Pauli operator in an original Hamiltonian should be
encoded to a certain multi-partite coupling term in an encoded Hamiltonian. Thus one needs to find
a scheme to implement such a term in the coherent computing model. It is highly nontrivial to
introduce, {\em e.g.}, a four-partite coupling among slave lasers. Further investigation is needed for
the usability of error correction codes.
Another scheme is dynamical decoupling. This scheme looks effective for suppressing thermal fluctuations
at a glance. Consider the minimum gap between two distinct eigenvalues of a problem Hamiltonian and normalize it with
the maximum gap. This decreases only polynomially in $n$ for any instance of the Ising spin configuration problem by
the definition of the problem. Thus the minimum operation interval of dynamical decoupling required for an effective
noise suppression decreases only polynomially in $n$ according to Eq. (52) of Ref.~\cite{Ng11}. One problem is how
to use this scheme for cancelling calibration errors. In addition, we need to find an implementation of the scheme
such that the scheme itself does not introduce an uncontrollable noise. This will be difficult for large $n$ because
imperfections in decoupling operations probably lead to a similar argument as Proposition~\ref{prop1}.

As we have proved, there are hard instances of the NPC Ising spin configuration problem for which one cannot efficiently
achieve a correct decision in the coherent computing model (Theorem~\ref{theo1}). This is a reasonable result in
light of the fact that no known conventional computer model could solve an NP-complete problem within a polynomial
cost. It is still an open problem if an unreasonable computational power is achievable by combining error protection
schemes with the coherent computing model.

\section{Conclusion}\label{secconclusion}
The model of coherent computing has been theoretically investigated in view of computational
cost under a realistic setting. It has been proved that there exist hard instances of the NPC Ising spin
configuration problem, which require exponential time for a correct decision in the model.

\subparagraph*{Acknowledgements}
The author would like to thank William J. Munro, Kae Nemoto, and Yoshihisa Yamamoto for helpful
discussions. This work is supported by the Grant-in-Aid for Scientific Research from JSPS
(Grant No. 25871052). 

%\appendix

%%
%% Bibliography
%%

%% Either use bibtex (recommended), but commented out in this sample

%\bibliography{dummybib}

%% .. or use bibitems explicitely

%\nocite{Simpson}

\end{document}